# Vibrations of Sessile Drops of Soft Hydrogels

Aditi Chakrabarti* and Manoj K. Chaudhury**
Department of Chemical Engineering
Lehigh University, Bethlehem, PA 18015
Phone: (+1) 610 758 4471

**Abstract.**

Sessile drops of soft hydrogels were vibrated vertically by subjecting them to a mechanically-induced Gaussian white noise. Power spectra of the surface fluctuation of the gel allowed identification of its resonant frequency that decreases with their mass, but increases with its shear modulus. The principal resonant frequencies of the spheroidal modes of the gel of shear moduli ranging from 55 Pa to 290 Pa were closest to the lowest Rayleigh mode of vibration of a drop of pure water. These observations coupled with the fact that the resonance frequency varies inversely as the square root of the mass in all cases suggest that they primarily correspond to the capillary (or a pseudo-capillary) mode of drop vibration. The contact angles of the gel drops also increase with the modulus of the gel. When the resonance frequencies are corrected for the wetting angles, and plotted against the fundamental frequency scale $(\gamma/m)^{0.5}$, all the data collapse nicely on a single plot provided that the latter is shifted by a shear modulus dependent factor $(1+\mu L/\gamma)$. A length scale $L$, independent of both the modulus and the mass of the drop emerges from such a fit.

*Email: adc312@lehigh.edu

** Email: mkc4@lehigh.edu

## 1. Introduction.

In recent years, estimation of the surface tension and the elasticity of soft gels has become the subject of considerable interests [1,2]. Starting with the pioneering study of Harden, Pleiner and Pincus (HPP) [3], several studies have focused on identifying the capillary and the elastic modes of vibration of either a half space or a thin film of the gel in terms of the various wave vectors that its free surface displays [4-7]. There is also a long history of studying the vibration modes of the surface of spherical liquid drops [8-17]. These studies start with the original prediction of Rayleigh [8,9], which show that the spherical harmonics of the capillary oscillations of an incompressible liquid drop surrounded by a rarified medium, are given as follows:

$$\omega_l = \sqrt{l(l-1)(l+2)\frac{\gamma}{\rho R^3}} \tag{1}$$

Where, $\omega_l$ is the resonant frequency, $l$ (=2,3,4…) is the eigen-mode of the oscillation, $R$ is its undeformed radius; $\gamma$ and $\rho$ are its surface tension and density respectively. For the case of a sessile drop, several authors have modified Rayleigh's equation with appropriate slip and no-slip boundary conditions at the three phase contact line [10-14]. Nevertheless, the basic scaling relation $\omega_l \sim R^{-3/2}$ has been found to be preserved in all the subsequent modifications of Rayleigh's equation.

One advantage of studying the spherical harmonics of a sessile liquid or gel drop is that the wave vector of the surface vibration is uniquely determined by its perimeter. In that spirit, we recently studied [2] the surface vibration of a soft (shear modulus, 40 Pa) hemispherical gel preformed on a flat substrate and found that the variation of its resonance mode as a function of its volume is very similar to that of pure hemispherical drop of water [15]. Based on these previous results, here we venture to investigate how these modes depend on the elasticity of the gel with a wide variation of its mass.

A simple energy analysis suggests that for a conservative system, the frequency of vibration would follow the following relationship:

$$\omega \sim \left(\frac{\gamma + \mu R}{\rho V}\right)^{1/2} \tag{2}$$

Where, $V$ is the volume of the gel sphere. Similar equation has been observed by various authors for the case of a half space, in which the surface wave vector takes the place of $1/R$ [1,3-6]. In our problem, as stated above, the wave vector is pinned by the perimeter of the drop. There are two extreme limits to the above equation. With a elasto-capillary number $\mu R/\gamma <<1$, only the capillary mode should prevail, in which case the frequency of vibration would vary inversely as the square root of the volume of the drop. On the other hand with $\mu R/\gamma >> 1$, the frequency of vibration would vary inversely as the cubic root of the volume. Both types of scaling have indeed been reported for the cases of pure liquid drops [15] as well for solid spheres of high elastic modulus [18]. In the intermediate range a pseudo-capillary mode with an effective surface tension $\gamma + \mu R$ should be in effect; however the frequency of vibration would depend on volume with an exponent lying between of -1/3 and -1/2.

The scaling relation as captured in equation (2), however, is valid for a conservative system. In the case with a real gel, we also have to consider the effect of surface and bulk viscous effects. One obvious consequence of viscosity is that the surface and the elastic modes may not be in phase; i.e. they may not necessarily coalesce as suggested by the simple scaling relation (2). In fact, for the cases of a surface of a vibrating half space of a gel, HPP found that there is a transition region when capillary and the elastic modes appear as a doublet, i.e. the two modes do not coalesce [3]. Even when these modes have a phase difference and they may not coalesce uniquely, there may still be an influence of one over the other. In that case, a pseudo capillary effect may manifest with an effective surface tension: $\gamma + \mu L$; but the length scale $L$ now may deviate from the radius of the drop $R$, the details of which would have to be deciphered from a viscosity dependent analysis of the phase difference of the capillary and elastic modes of a vibrating drop.

At this juncture, we should point out a couple of other reasons why a simple relation of the type shown in equation (2) may not be valid with a real gel. The surface tension term that appears in equation (2) could actually be the surface stress [19] given as $\gamma + d\gamma/d\varepsilon$, $\varepsilon$ being the surface strain. Surface rheology may have a frequency spectrum that is different from that of the bulk deformation. Thus, not only the surface tension of a vibrating gel could be different from the thermodynamic surface excess free energy, but also there could be other non-trivial reasons why there could be a phase difference between surface and bulk modes of vibration. Furthermore, both the rheological spectra could also be non-linear in frequency that may not be captured with a simple Maxwell-like viscoelastic model.

In view of the above mentioned complexities, our objectives of this study are quite modest. First, we intend to find out if the spheroidal mode of vibration ($l=2$) could be detected for soft gel drops of different elastic moduli and how the resonance frequency varies with the volume of the sphere. Secondly, we wish to find out if the observed frequency could be expressed in terms of the surface tension and the shear modulus of the gel in a simple semi-empirical form.

In order to achieve this objective, we developed a new method to produce gels of different radii, which are as spherical as possible. We then placed these gels on a hydrophobic substrate on which the gels subtend well-defined contact angles > 90°. The gels are then vibrated vertically using a mechanically induced Gaussian random noise, the power spectra of which helped us identify the resonance frequency corresponding to the $l=2$ mode. In the subsequent sections, we first describe the experimental method of preparing such spherical gels and the method used to identify their spheroidal modes. Followed by the experimental section, we analyze how these modes depend on the volume and shear modulus (55 Pa to 290 Pa) of the gel and show how the data can be collapsed about a single line with an appropriate scaling relation.

## 2. Materials and Methods

### 2.1. Measurement of Elastic Moduli of Gels

Before performing the vibration experiments with the spherical drops of the gel, we needed to estimate their shear moduli independently. The shear moduli of the different gels were determined by vibrating a thin slab of a gel in the shear mode in a confined geometry, the details of which are explained in our previous publications [1,20]. Here, we briefly describe the main idea behind such a measurement. A schematic of the experimental setup is shown in Figure 1a (More details can be found in Ref. [20]). A thin slab of gel was cured in between two clean glass plates (Fisherbrand, top: 25mm x 75mm, 1mm; lower: 50mm x 75mm, 1mm) separated by uniform spacers. Once the gel was cured, a steel disk was fixed to the upper glass plate of the sample. The lower plate of the sample was fixed atop a platform mounted on a vibration isolation table (Micro-g, TMC). A strong magnet was fixed at the edge of an aluminum stage, which in turn was connected to a mechanical oscillator (Pasco Scientific, Model no. SF-9324). Random white noise was generated using a waveform generator (Agilent, model 33120A) that was transferred to the mechanical oscillator and thus generating a random magnetic field. When the magnet was brought close enough to the gel sample, the thin gel slab vibrated randomly in a shear mode due to the random vibration of the steel disk attached to the upper plate. A high speed movie (Redlake Motion Pro, model no 2000, at 1000frames/s) capturing the random motion of the gel slab and its subsequent analysis in a motion tracking software (Midas 2.0, Xcitex Inc., USA) yielded the displacement fluctuations. The resonant peaks (Figure 1b) of the shear vibrations were identified from the power spectra of these fluctuations by averaging data taken from ten videos each had a duration of 4s. The shear modulus was estimated by using the formula $\mu = 4\pi^2 \omega^2 mH/A$, where $\omega$ is the resonant frequency, $m$ is the mass of the glass plate and the steel disk, $H$ is the thickness of the gel slab and $A$ is the area of contact of the gel slab with the glass plates.

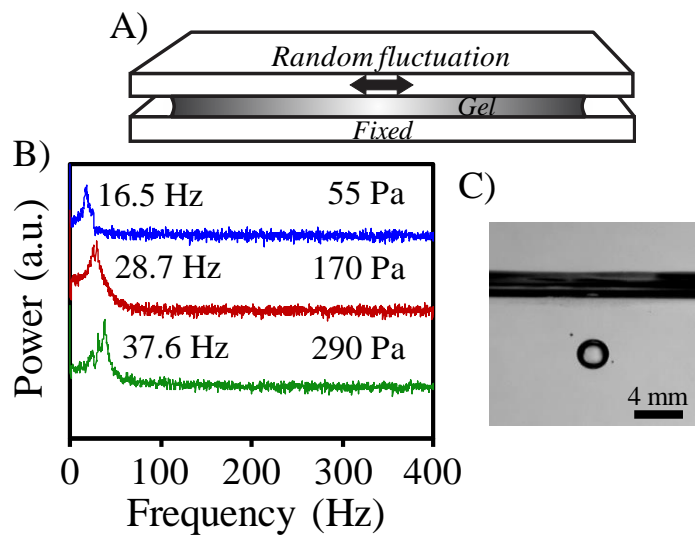

**Figure 1.** A) Schematic of a thin gel slab (1 mm thick) sandwiched between two parallel glass plates undergoing shear vibration. B) The resonant peaks of the shear vibration of gel as obtained from the power spectra of their random vibration (RMS fluctuation 0.03 mm). Powers are in logarithmic scale. C) An example of a spherical gel drop of shear modulus 55 Pa; this is a snapshot of a video that was captured while the sphere sank slowly through mineral oil in a quartz cell.

## 2.2. Preparation of Gel Spheres

Highly spherical gel spheres (Figure 1c) were made by curing polyacrylamide gel solution drops of different volumes by suspending them in a liquid density gradient. Three different gels (55 Pa,

170 Pa and 290 Pa) were used for the study, the details for the preparation of which are described elsewhere [1,20]. The density gradient was formed in small beakers with a liquid heavier than the gel solution (PDM-7040, Gelest Inc., density 1.07 g/cc) at the bottom and a lighter liquid (n-octane, 97% pure, Acros organics, density 0.7 g/cc) on the top. After all the ingredients of the gel were mixed, different volumes of the gel solution were released gently over the top surface of the octane in the container housing the density gradient. The drops of gel solution become neutrally buoyant in between the two liquids forming density gradient. These suspended gel drops cure to form highly spherical gel drops (Figure 1c) that were subsequently washed in fresh n-heptane (Fisher Chemicals) repeatedly and dried moderately in air. The volume of gel spheres thus formed ranged from $2\mu L$ to $100\mu L$ as determined from their weights.

### 2.3. Vibration Studies of Gel Spheres

A rectangular polystyrene cuvette was used to house each gel sphere for the vibration studies (Figure 2a). A small glass piece (8 mm x 8 mm, 1mm), was hydrophobized using the usual method of reacting with dodecyltrichlorosilane (Gelest Inc.) and was fixed on one of the walls inside the polystyrene cuvette. As soon as the gel sphere was placed on the hydrophobic glass piece, the open end was sealed firmly with parafilm to avoid drying of the gel during the time the studies were carried out. Filter paper soaked with deionized water stacked inside the cuvette maintained its relative humidity to about 99.9%. The cuvette housing the sample was then fixed onto an aluminum stage that was set to vertical vibration with random white noise by the mechanical oscillator and waveform generator as described in section 2.1. The surface fluctuations of the gel drops were video recorded with a high speed camera at 1000 frames/s that were subsequently fast Fourier transformed to yield their power spectra (Figure 2b). The resonant frequency for the spheroidal mode was identified from the power spectrum for each gel sphere. As we reported earlier [15], the probability distribution function of the accelerations of the stage was Gaussian that was white up to a practical bandwidth of 5 kHz. With these random accelerations ($\gamma$), the strength of the noise was estimated as, $K=\langle \gamma^2(t) \rangle \tau_c$, where $\gamma(t)$ is the value ($m/s^2$) of the noise pulse and $\tau_c$ is its duration (40 $\mu s$). A constant noise strength (0.1 $m^2/s^3$) was used to perform all the vibration studies although we noted that the primary peak position for each spheroidal vibration mode is independent of the noise strength (Figure 2b). Although an effective temperature can be obtained by multiplying $K$ with the mass of the object and its characteristic relaxation time, we refrained from doing so in this work. Here we present directly the root mean square value of the surface fluctuation to distinguish the states of the system.

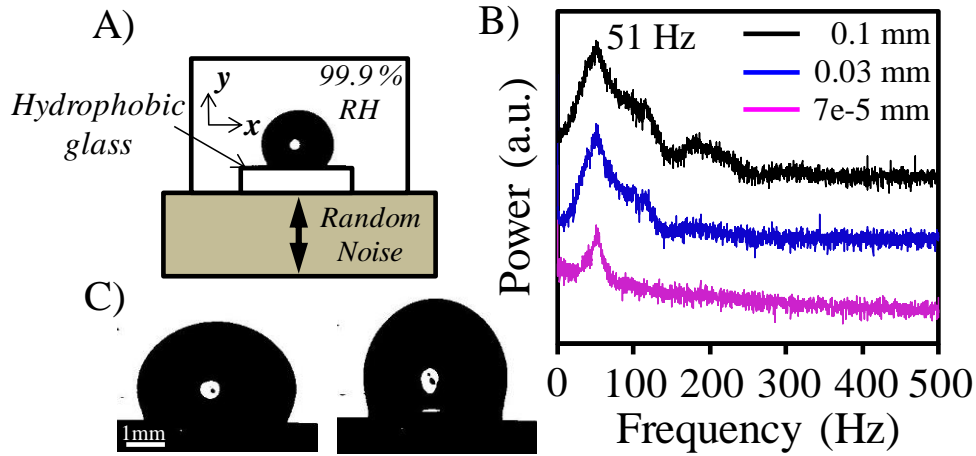

**Figure 2.** A) Schematic of the experimental setup for studying the height fluctuation of the gel drops placed on a hydrophobic substrate after subjecting it to a random white noise. B) The power spectra for a 38$\mu$L gel sphere (55 Pa) at different noise strengths, each has its resonant mode corresponding to $l=2$ at 51 Hz, their RMS fluctuation being marked in legend. C) Two randomly selected snapshots of vibration of a 47$\mu$L (55 Pa) gel sphere from a high speed movie of it undergoing random fluctuation. This corresponds to a spheroidal mode of $l=2$.

## 3. Experimental Results and Discussion

The resonance frequency of the spheroidal mode ($l=2$) of each gel drop was identified from the power spectrum of its surface fluctuation (Figure 3a). For gels of modulus spanning by a factor of six, the resonance frequencies varied inversely with volume with an exponent close to 0.5, which are not too different from the previously published values of pure hemispherical water drops [15]. The frequencies for these gels of different moduli (Figure 3a) do not diverge significantly with the increase of the size of the gel sphere nor do they converge at small drop sizes.

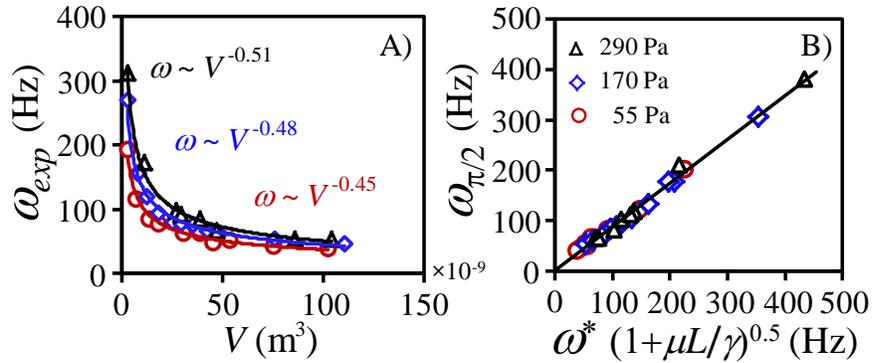

**Figure 3.** A) Experimentally observed resonant frequencies for the spheroidal mode ($l=2$) of vibration of gel drops of different shear moduli plotted as a function of their volume. B) All the resonance frequencies are shifted to a contact angle of 90° plotted as a function of the frequency $\omega^* = \sqrt{\gamma/m}$ multiplied by a dimensionless function $\sqrt{1+\mu L/\gamma}$. Here $L$ is the only adjustable parameter. For $L=1.5$mm, the best linear fit has a slope ~0.9 with a regression coefficient of 0.99.

These observations contrast Equation (2) that suggests that the resolution of the frequencies should increase with the drop volume as the modulus increases and they would approach each other at small volume as the elastocapillary number decreases.

The contact angle of the gel drops on the hydrophobic substrate, however, increase with its modulus: ($\theta =100°$ for 55 Pa, 115° for 170 Pa and 125° for 290 Pa). Since the contact angle influences the perimeter of the drop and thus the wave vector, all the experimentally observed frequencies were shifted to that of $\theta=90°$ ($\omega_{\pi/2}$) using the following scaling factor [12]:

$$\omega_\theta = f(\theta)\ \omega^* \qquad (3)$$

Where, $\omega^* = \sqrt{\gamma/m}$ and $f(\theta) = \left[\dfrac{2.78(2 - 3\cos\theta + \cos^3\theta)}{\theta^3}\tanh\left(\dfrac{\pi}{2\theta}(2 - 3\cos\theta + \cos^3\theta)\right)\right]^{1/2}$

Where, $m$ is the mass of the gel drop, the contact line of which essentially pinned on the substrate during vibration. These scaled frequencies ($\omega_{\pi/2}$) obtained for gels of different shear moduli (55 Pa- 290 Pa) could be collapsed (Figure 3b) nicely about a single line when the fundamental capillary frequency scale $(\gamma/m)^{0.5}$ is shifted as follows:

$$\omega_{\pi/2} \sim \omega^*\sqrt{1 + \mu L/\gamma} \qquad (4)$$

With the value of $L$ set at 1.5 mm for all the gels, the data $\omega_{\pi/2}$ vs $\omega^*\sqrt{1 + \mu L/\gamma}$ collapsed nicely about a straight line (regression coefficient of 0.99), with a slope (0.9) that is very close to that (1.0) of a similar plot for pure water drops (where $\mu = 0$) as published previously [15]. Some additional remarks are in order here. Firstly, we noted that the resonant mode corresponding to $l=2$ is rather insensitive to the strength of the noise (Figure 2b). This coupled with the observation that the random displacement of the surface is Gaussian (Figure 4) suggest that these gels behave more or less linearly as was also observed previously with pure water drops [15].

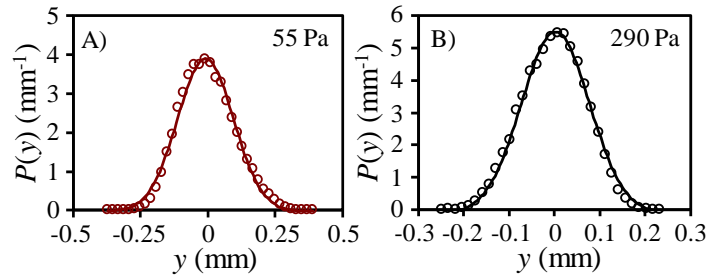

**Figure 4.** A-B) Examples of the height fluctuations of a $38\mu L$ gel sphere of two different moduli (55 Pa and 290 Pa) following a Gaussian probability distribution. Noise strength is 0.1 m$^2$/s$^3$.

Although the primary mode of vibration does not depend on the noise, another mode adjacent to the primary mode becomes pronounced with the increase of the noise strength (Figure 2b). From our previous observations with water drops, we feel that this adjacent mode does not correspond to the $l=3$ capillary mode, which would appear at a significantly higher frequency. Most likely, this is an elastic mode which is sensitive to the amplitude of vibration. If this is so, we are probably observing a doublet as was envisaged by HPP for a semi-infinite gel in the transition region [3].

Although the origin of the value of $L$ being independent of the physical parameters of the gels remains elusive at present, it poses several interesting scenarios. One scenario could be that a thin film region exists on the surface of the gel, the elastic strain energy of which adds to the conventional surface energy ($\gamma$). The value of $L$ (1.5 mm) being comparable to the elasticity modified Laplace length [$(\sqrt{\gamma/\rho g})\exp(-B\mu)$] as reported in Ref. [20] of these gels is also striking. Is $L$ then related to the elasto-capillary number, i.e. is closer to the height of the drop that varies slower than its radius as the volume of the drop increases ? We are not sure at present. A detailed analysis of the vibration modes of the gel along the line of HPP, but in polar spherical coordinate, is critical to make further inroads to the problem. The statistical mechanical aspect of the current problem is similar as well as different from that studied by HPP in that these authors consider the roles of thermal fluctuations in which the autocorrelation of random stress is related to temperature and viscosity via usual fluctuation dissipation relation. In our problem both the thermal and an externally imposed random mechanical noises are in effect, in which the autocorrelation of random stress would depend on both thermodynamic temperature as well as the strength of the mechanical noise manifesting in terms of an "effective temperature". The Gaussian nature of the displacement fluctuations suggest that viscous term enter in the problem linearly, which should simplify the analysis. The problem is challenging, but it also provides opportunities for new physics to be discovered as the externally imposed noise and internal friction are de-coupled, and thus the role of friction could be studied independently of noise, which is not the case with thermal systems. These studies, when carried out in conjunction with the analysis of the vibration of a half space should also enhance the scope of these studies.

## 4. Summarizing Comments

What we presented here is a report of a preliminary experimental study, in which the spheroidal ($l=2$) mode of vibration of a sessile soft gel drop is studied as a function of its volume and elasticity. The resonance mode is found to vary with volume as is the case with a liquid drop governed by capillarity, but its effective surface tension is weakly dependent on the modulus as expected of a pseudo-capillary mode. By replacing the radius of the drop with a fixed length scale $L$, all the resonance frequencies could be collapsed about a line obtained from a plot of $\omega_{\pi/2}$ vs $\omega^*\sqrt{1+\mu L/\gamma}$. Admittedly, the physics behind this scaling is not clear to us. We have not taken into account the detailed viscoelastic properties of the gels, although the sharpness of the shear deformation peaks (figure 1) suggest that viscous damping in these gels is probably not very large. The low viscous damping in these films were also evident in another measurement (to be published separately) in which a silanized glass disc was pulled off a gel film (150 to 200 µm thick) bonded to a rigid substrate at different velocities. Adhesive separation occurred at the disk/gel interface. For gels of moduli ranging from 40 to 330 Pa, the adhesive pull-off stress varied only a factor of three within three decades pull-off velocity. Since adhesion tests are very sensitive to viscous dissipation in a soft adhesive, the lack of significant dependence of adhesive fracture stress on speed suggests that these gels are mainly elastic. Nevertheless, detailed rheological measurements are necessary for making further progress in the interpretations of these results, which is a subject of our future study. In this regard, the beautiful technique developed by Pottier et al [21,22], which allows measurement of surface rheological properties of a soft object from the thermal fluctuation spectrum will be ideally suited for further understanding the roles of elasto-capillarity and (possible) viscous damping on the eigen-modes of the spherical drops as reported here. Nevertheless, the simplicity of the equation $\omega \sim [(\gamma + \mu L)/\rho V]^{1/2}$ with one adjustable parameter (*L*) that successfully collapsed all the data obtained with various gel spheres on one master curve motivates us to investigate in future why only one constant value of *L* was needed and what is its physical origin. We hope that this preliminary study will motivate further experimental and theoretical studies of the vibration of spherical gels. Presently, we have investigated only the spheroidal mode of vibration corresponding to $l$=2 for the gel spheres. We expect that more in-depth insights into the gel vibration problem can be gained by studying its higher modes, which are activated with increasing noise strengths. These detailed studies are reserve for future.

**Acknowledgements**

We thank Dr. Christian Fretigny (ESPCI) for some valuable discussions, especially for suggesting that the length scale *L* may be related to the height of the drop, hence the elasto-capillary length.